\documentclass[aps,prd,10pt,twocolumn,superscriptaddress,floatfix,nofootinbib,amsmath,amssymb,altaffilletter,preprintnumbers]{revtex4-2}
\usepackage[colorlinks=true,pdfstartview=FitV]{hyperref}

\usepackage{orcidlink}
\usepackage{graphicx}
\usepackage{hyperref}
\hypersetup{allcolors=[rgb]{0.0 0.0 1.0},linkcolor=[rgb]{0.75 0.05 0.05},colorlinks=true,breaklinks=true}
\usepackage{amsmath}
\usepackage{amssymb}
\usepackage{bbold}
\usepackage{multirow}
\usepackage{bm}
\usepackage{hyperref}
\long\def\exclude#1{}
\usepackage{multirow}
\usepackage{soul}
\usepackage{natbib}
\usepackage{amsmath}
\usepackage{fontawesome}
\usepackage{mathrsfs}
\usepackage{amssymb}
\usepackage{bm}
\usepackage{yfonts}
\usepackage{makecell}
\usepackage{color}
\usepackage{xspace}
\usepackage{url}
\usepackage{verbatim}
\usepackage{mathtools}
\usepackage{upgreek}
\usepackage{units}
\usepackage{soul}
\usepackage{amstext}
\usepackage[sc]{mathpazo}
\usepackage{helvet}
\usepackage{booktabs}
\usepackage{gensymb}
\usepackage{tabulary}
\usepackage{tabularx}
\usepackage{etoolbox}
\usepackage[version=4]{mhchem}
\usepackage[utf8]{inputenc}
\newcolumntype{Y}{>{\centering\arraybackslash}X}
\usepackage[caption=false]{subfig}
\usepackage[official]{eurosym}
\definecolor{lightgray}{rgb}{0.9,0.9,0.9}	    
\definecolor{green}{rgb}{0,0.5,0}
\definecolor{red}{rgb}{1,0,0}
\definecolor{blue}{rgb}{0,0,0.5}

\long\def\symbolfootnote[#1]#2{\begingroup%
\def\thefootnote{\fnsymbol{footnote}}\footnotetext[#1]{#2}\footnotemark[#1]\endgroup}

\newcommand{\dbd}[2]{\ifmmode \frac{\textrm{d}#1}{\textrm{d}#2}\else $\textrm{d}#1/\textrm{d}#2$\fi}
\newcommand{\pbp}[2]{\ifmmode \frac{\partial#1}{\partial#2}\else $\partial#1/\partial#2$\fi}

\DeclareMathAlphabet{\mathpzc}{OT1}{pzc}{m}{it}
 \newcommand{\eV}{\text{e\kern-0.15ex V}\xspace}

 \newcommand{\TeV}{\text{T\kern-0.1ex \eV}\xspace}

\DeclareMathAlphabet{\mathpzc}{OT1}{pzc}{m}{it}

\newcommand{\be}{\begin{equation}}
\newcommand{\ee}{\end{equation}}
\newcommand{\bea}{\begin{eqnarray}}
\newcommand{\eea}{\end{eqnarray}}

\renewcommand\({\left(}
\renewcommand\){\right)}

\begin{abstract}
\noindent
Searches for high frequency gravitational waves using cavities based on the Gertsenshtein effect were recently proposed, building off existing axion dark matter experiments. In particular, the sensitivity of axion dark matter experiments using metamaterial plasmas (tunable plasma haloscopes) to gravitational waves has not been explored in detail. Here we perform a full analysis of gravitational wave detection in plasma haloscopes, showing that the baseline design of experiments such as ALPHA is several orders of magnitude less sensitive than previously thought. We show how simple changes to the experiment can recover that sensitivity and lead to a powerful gravitational wave detector in the ${\cal O}(10-50)$~GHz frequency range.
\end{abstract}

\begin{document}

\preprint{FERMILAB-PUB-24-0436-T}

\title{Gravitational Wave Detection With Plasma Haloscopes}

\author{Rodolfo Capdevilla
\orcidlink{0000-0002-0122-7704}}
\affiliation{Theory Division, Fermi National Accelerator Laboratory, Batavia, IL 60510, USA}

\author{Graciela B. Gelmini
\orcidlink{0000-0001-8401-1505}}
\affiliation{Department of Physics and Astronomy, UCLA, 475 Portola Plaza, Los Angeles, CA 90095, USA}

\author{Jonah Hyman~\orcidlink{0009-0007-5575-9128}} 
\affiliation{Department of Physics and Astronomy, UCLA, 475 Portola Plaza, Los Angeles, CA 90095, USA}

\author{Alexander J. Millar~\orcidlink{0000-0003-3526-0526}}
\affiliation{Theoretical Physics Division, Fermi National Accelerator Laboratory, Batavia, IL 60510, USA}
\affiliation{Superconducting Quantum Materials and Systems Center (SQMS), Fermi National Accelerator Laboratory, Batavia, IL 60510, USA}

\author{Edoardo Vitagliano
\orcidlink{0000-0001-7847-1281}}
\affiliation{Dipartimento di Fisica e Astronomia,
Università degli Studi di Padova, Via Marzolo 8, 35131 Padova, Italy}
\affiliation{Istituto Nazionale di Fisica Nucleare (INFN),
Sezione di Padova, Via Marzolo 8, 35131 Padova, Italy}

\maketitle

\section{Introduction}

In the last decade gravitational waves (GWs) 
have become a crucial astrophysical probe, complementing cosmic ray, electromagnetic, and neutrino observations. Today, a series of experiments observe the GW sky and others are planned to further
advance the possibilities of multi-messenger astronomy. Although astronomy based on photons~\cite{Ressell:1989rz,Hill:2018trh} and neutrinos~\cite{Vitagliano:2019yzm} can rely on signals with frequencies that largely exceed the kilohertz range, the focus of existing and forthcoming GW detectors is on frequencies ranging approximately from $10^{-16}$ to $10^3$~Hz~\cite{Moore:2014lga,LIGOScientific:2017vwq,LIGOScientific:2016aoc,EPTA:2023fyk,Reardon:2023gzh,NANOGrav:2023gor,Xu:2023wog}. At the lowest part of this range,
Cosmic Microwave Background probes can constrain the stochastic fHz GW background radiation, arising e.g. from inflation~\cite{BICEP2:2015nss,BICEP2:2018kqh,Planck:2018vyg} (see Ref.~\cite{Achucarro:2022qrl} for a review). At larger frequencies ($>10^{-4}$ Hz) black hole, neutron star and mixed binaries are expected to contribute to the GW sky, with a maximum frequency that can be roughly estimated as $\omega\simeq \sqrt{GM/R^3}\simeq \mathrm{kHz}(M_\odot/M)$ (where $G$ is the Newton constant, $M$ is the reduced mass 
of the binary, $M_\odot$ is a solar mass and $R$ is the innermost stable circular orbit)~\cite{Maggiore:2018sht}. Pulsar 
timing arrays are employed to detect a signal at nHz frequencies~\cite{InternationalPulsarTimingArray:2023mzf}, 
possibly produced by supermassive black 
hole binaries 
~\cite{Begelman:1980vb}. At yet larger frequencies, interferometers can 
detect kHz GWs produced by inspiraling compact binaries of stellar remnants ($M\simeq \mathcal{O}(1-100) \, M_\odot$)~\cite{LIGOScientific:2016aoc,LIGOScientific:2017vwq}. Primordial black holes (PBHs) formed in the early Universe prior to any galaxies and stars are 
viable dark matter candidates, 
and might even constitute 100\% of the dark matter if they are in the
``asteroid-mass" range, 
$10^{-16}$ to $10^{-10}\, M_\odot$~\cite{Carr:2020gox}. The merging of such objects would produce high-frequency GWs, which could provide a possible probe of asteroid-mass PBHs~\cite{Dolgov:2011cq, Franciolini:2022htd,Profumo:2024okx}. 

The possibility of detecting high frequency GWs (HFGW), roughly 10$^{-2}$ to 10$^4$ MHz, through graviton-photon conversion (the inverse-Gertsenshtein effect~\cite{Gertsenshtein:1961xxx,1974JETP...38..652Z}) has been explored in recent years for various types of axion haloscopes~\cite{Ballantini:2003nt,Ito:2019wcb,Ejlli:2019bqj,Berlin:2021txa,Domcke:2022rgu,Berlin:2023grv,Bringmann:2023gba,Domcke:2023bat,Kahn:2023mrj,Navarro:2023eii,Gatti:2024mde,Domcke:2024mfu,Domcke:2024eti} (see also~\cite{Carney:2023nzz,Ahn:2023mrg,Ratzinger:2024spd}). 
These GWs would necessarily be of astrophysical origin because to be within reach of these detectors they should be so intense that if they were of cosmological origin their density would be $\Omega_{\rm GW} h^2 \gtrsim 10^{-6}$, above the BBN and CMB upper limit originating from bounds on the effective number of neutrino species $N_{\rm eff}$~\cite{Pagano:2015hma}.
Possible astrophysical sources of these GWs include mergers of sub-solar mass objects (e.g. PBHs or other exotic objects)~\cite{Giudice:2016zpa}, and the annihilation of bosons in clouds produced by black hole superradiance~\cite{Arvanitaki:2012cn} (see also e.g.~\cite{Aggarwal:2020umq,Aggarwal:2020olq} and references therein). 

Here we explore the graviton-photon conversion within a resonant cavity, not empty as done in~Ref.~\cite{Berlin:2021txa}, but rather filled with a tunable plasma, in which case the resonance frequencies are shifted by the plasma frequency $\omega_{\rm p}$. Tunable axion haloscopes consisting of wire metamaterial were proposed as axion~\cite{Lawson:2019brd} and dark photon~\cite{Gelmini:2020kcu} detectors, and prototypes are being developed by the Axion Longitudinal Plasma HAloscope (ALPHA) Collaboration~\cite{ALPHA:2022rxj}. The separation of parallel wires in these detectors allows one to change $\omega_{\rm p}$ and thus
the resonant frequencies of the detector. Previous studies have estimated the sensitivity of tunable plasma to GWs, focusing primarily on analyzing the quality and coherence of potential sources \cite{Gatti:2024mde} (see also~\cite{Alesini:2023qed} for a similar discussion). In contrast, this work emphasizes a more precise estimate of ALPHA's sensitivity. We analyze the proposed ALPHA configuration, where the tunable plasma is fully anisotropic, and explore the case of an isotropic setup, finding that the latter is actually more sensitive to HFGWs.

The paper is structured as follows. In Section~\ref{sec:analyticalformalism} we start by developing the analytic formalism describing the interaction between a GW and a resonant detector. We show how medium effects can be taken into account in Section~\ref{sec:mediumeffects}, and present the prospects for GW detection in a plasma haloscope, such as ALPHA, in Section~\ref{sec:projections}. Section~\ref{sec:conclusions} is devoted to
a summary and discussion.

\section{Analytical Formalism}\label{sec:analyticalformalism}

In this section, we provide the theoretical ingredients needed for our analysis. We want to describe in the language of classical fields the interaction between a propagating gravitational perturbation, i.e., a GW, and a stationary electromagnetic system in the laboratory, i.e., a plasma haloscope (PH). Our starting point is the action for the electromagnetic potential $A^\mu$ in the presence of a current density $J^\nu$ in curved space-time. Once we consider linear perturbations of the metric, we define a new gravitationally induced current density $J^\nu_{\rm eff}$ which encodes the effect of gravitational perturbations on electromagnetic systems. In order to describe the interaction between the GW and the PH, we need to define the former in the Proper Detector (PD) frame.

\subsection{Lagrangian and Proper Detector Frame}
\label{ss.LandPD}

Maxwell's equations in curved space-time can be derived from the Maxwell-Einstein action
\be
S=\int dx \sqrt{-g}\left( -\frac{1}{4}g^{\mu\nu}g^{\alpha\beta} F_{\mu\alpha}F_{\nu\beta} + g^{\mu\nu}A_\mu J_\nu \right),
\ee
where $F_{\mu\nu}=\partial_\mu A_\nu-\partial_\nu A_\mu$ is the electromagnetic field  tensor,  $g^{\mu\nu}$ is the space-time metric tensor, and $g={\det}(g_{\mu\nu})$. In linearized gravity we write $g^{\mu\nu}=\eta^{\mu\nu}-h^{\mu\nu}$, where $\eta^{\mu\nu}={\rm diag}(-1,1,1,1)$ and $|h^{\mu\nu}|\ll 1$ is the dimensionless strain and represents a small gravitational perturbation on the flat metric, so one can neglect terms that are quadratic in  $h^{\mu\nu}$. The action above contains the following terms,
\be
S\supset\int dx \left( \mathcal{L}_0 +\frac{h^{\lambda\sigma}}{2}P_{\mu\nu\lambda\alpha\sigma\beta}F^{\mu\alpha}F^{\nu\beta} - h^{\alpha\beta}P_{\alpha\mu\beta\nu}A^\mu J^\nu \right),
\ee
with 
\begin{align}
\mathcal{L}_0 & = -\frac{1}{4} F^{\mu\nu}F_{\mu\nu} + A^\mu J_\mu,\\
P_{\mu\nu\lambda\alpha\sigma\beta} & = \eta_{\mu\nu}\eta_{\lambda\alpha}\eta_{\sigma\beta}-\frac{1}{4}\eta_{\lambda\sigma}\eta_{\mu\nu}\eta_{\alpha\beta},\\
P_{\alpha\mu\beta\nu} & = \eta_{\alpha\mu}\eta_{\beta\nu}.
\end{align}
Varying the action with respect to the field $A^\mu$ one can obtain the inhomogeneous Maxwell's equations
\be
\partial_\mu F^{\mu\nu} = -J^\nu - J^\nu_{\rm eff},
\label{eq:Maxwell_covariant}
\ee
where we have defined a gravitationally-induced effective current
\be
J^\nu_{\rm eff} = -h^{\nu\alpha}J_\alpha +\frac{h}{2} J^\nu - \partial_\mu \left( h^{\nu\alpha}F^\mu_{\,\,\,\alpha} - h^{\mu\alpha}F^\nu_{\,\,\,\alpha} - \frac{h}{2}F^{\mu\nu} \right).
\label{eq:Jeff}
\ee
Here, $h \equiv h^{\mu}_{\,\,\mu}$. Conceptually, this current emerges from the effect of linear gravitational perturbations on electromagnetic systems. When the GW overlaps with the PH, due to the coupling between the strain and the electromagnetic field tensor, in presence of an external magnetic field the gravitational perturbation turns into an electromagnetic signal that acts as a driving force on the PH. As we will see later, when the frequency of the GW matches the frequency of a particular mode of the PH, this effect is resonant~\cite{Berlin:2021txa}. In the absence of currents in the PH, i.e., setting $J^\mu=0$ in Eq.~\eqref{eq:Jeff}, the mechanism through which the gravitational signal turns into an electromagnetic signal is the Gertsenshtein effect~\cite{Gertsenshtein:1961xxx,1974JETP...38..652Z}.

To describe the GW during its overlap with the cavity, we need the explicit form of the strain $h_{\mu\nu}$. This tensor has a very simple form in the transverse-traceless (TT) frame, where it satisfies $h_{0\mu}=0$, $h=\eta^{\mu\nu}h_{\mu\nu}=0$, and $\partial^\mu h_{\mu\nu}=0$. In the TT-frame, for a GW moving along the $z$ axis $h_{\mu\nu}$ has a simple plane wave-looking form
\begin{equation}
   h_{\mu\nu}^{\rm TT}= \left(\begin{array}{cccc}
0 & 0 & 0 & 0\\
0 & h_{+} & h_{\times} & 0\\
0 & h_{\times} & -h_{+} & 0\\
0 & 0 & 0 & 0
\end{array}\right)
e^{-i(t-z)\omega_g},
\end{equation}
where $\omega_g$ is the frequency of the GW, $+/\times$ correspond to the polarization of the GW, and $h_{+/\times}$ is the corresponding characteristic strain. For a GW moving along any direction described by the azimuthal and polar angles $(\varphi_g,\theta_g)$ the strain tensor in the TT-frame is
\begin{align}
h_{ij}^{\rm TT} & = \tilde{h}_{ij}^{\rm TT} e^{ik_g\cdot x},\\
\tilde{h}_{ij}^{\rm TT} & = \frac{1}{\sqrt{2}}\left[ (U_i U_j-V_iV_j)h_+ +(U_iV_j+V_iU_j)h_\times \right],
\end{align}
where $k_g=\omega_g(1,\hat{\bf k}_g)$ is the GW four-vector, and the unit vectors $\hat{\bf k}_g$, $\hat U$, and $\hat V$ are given by
\begin{align}
\hat{\bf k}_g & = (\sin\theta_g \cos\varphi_g,\sin\theta_g \sin\varphi_g,\cos\theta_g)\\
\hat U & = (\cos\theta_g \cos\varphi_g,\cos\theta_g \sin\varphi_g,-\sin\theta_g)\\
\hat V & = (-\sin\varphi_g,\cos\varphi_g,0).
\end{align}
In the TT frame, the coordinate system is attached to a free-falling test mass. This means that the description of the PH fixed to Earth's surface
is complicated in this frame. For this reason it is more appropriate to describe $h_{\mu\nu}$ in what is called the proper detector (PD) frame. This frame is defined by the rigid coordinates of the experiment fixed to the center of mass of the PH. In the PD frame, $h_{\mu\nu}$ is~\cite{Berlin:2021txa,Domcke:2022rgu}
\begin{align}
    h_{00} &= \omega_g^2 F({\bf k}_g\cdot {\bf r}) {\bf b}\cdot {\bf r}\\
    h_{0i} &= \frac{\omega_g^2}{2} \bigl[ F({\bf k}_g\cdot {\bf r})-i F'({\bf k}_g\cdot {\bf r})  \bigl] \nonumber\\
&\phantom{=\frac{\omega_g^2}{2} \bigl[ F({\bf k}_g\cdot {\bf r}) -} \times\bigl[ (\hat{\bf k}_g \cdot {\bf r})b_i-({\bf b}\cdot {\bf r})\hat{\bf k}_{g,i} \bigl], \\
    h_{ij} &= -i\omega_g^2F'({\bf k}_g\cdot {\bf r}) \bigl[ |{\bf r}|^2 h_{ij}^{\rm TT}({\bf r}=0) \nonumber\\
&\phantom{=i\omega_g^2F'({\bf k}_g\cdot {\bf r}) \bigl[} + ({\bf b}\cdot {\bf r}) \delta_{ij} - (b_ir_j+b_jr_i) \bigl],
\end{align}
where $F'$ is the derivative of $F$ with respect to its argument, and the explicit form of $F$ and ${\bf b}$ are
\begin{align}
    F(\xi) & = \frac{1}{\xi^2}(e^{i\xi}-1-i\xi),\\
    b_j & = r_i h_{ij}^{\rm TT}({\bf r}=0).
\end{align}

\subsection{Electric Field Signal}

From Eq.~\eqref{eq:Maxwell_covariant} we can write Maxwell's equations in the form
\begin{align}
{\bm\nabla}\cdot {\bf D} & = \rho_f + \rho_{\rm eff},\label{eq.Max1}\\
{\bm\nabla}\times {\bf B} - \bf{\dot{ D}} & = {\bf j}_f + {\bf j}_{\rm eff},\label{eq.Max2}\\
{\bm\nabla}\cdot {\bf B} & = 0,\label{eq.Max3}\\
{\bm\nabla}\times {\bf E} + \bf{\dot {B}} & = 0\label{eq.Max4},
\end{align}
where $\rho_f$ and ${\bf j}_f$ are the free charges and currents and $J^\mu_{\rm eff}=(\rho_{\rm eff}, {\bf j}_{\rm eff})$ defined in Eq.~(\ref{eq:Jeff}) contains all the information of the incident GW.
Here $D^i=\epsilon^{ij}  E_j$, $\epsilon$ being the electric permittivity tensor. We will assume that the medium is not magnetic (i.e., $\mu=1$). Combining Eqs.~\eqref{eq.Max2}~and~\eqref{eq.Max4} we obtain
\be
{\bm\nabla} \times {\bm\nabla} \times {\bf E} + \partial_t^2 {\bf D} = - \partial_t {\bf j}_f - \partial_t {\bf j}_{\rm eff}.
\label{eq.main_eq_0}
\ee
We will consider the PH to be a single moded cavity with a dielectric medium (i.e., the plasma) inside.  As noted in Refs.~\cite{Balafendiev:2022wua,ALPHA:2022rxj} this description holds as long as the decay length of the plasma in the medium ($\sqrt\epsilon/\omega$) is larger than the size of the cavity. In Appendix~\ref{app.modes} we provide more details on how the medium modifies the modes of the system with respect to those of an empty cavity for the anisotropic case, in which the PH metamaterial has wires only in one direction (the $z$-axis) and the isotropic case, in which wires are along all three perpendicular axes.
 Equation~\eqref{eq.main_eq_0} will allow us to describe the effect of the GW on the PH modes. For this purpose, it is customary to express the E-field inside the PH in terms of the solenoidal (${\bf E}_s$) and irrotational (${\bf E}_i$) modes, defined by
\begin{subequations}
\begin{align}
{\bm\nabla}\cdot {\bf E}_{s} & = 0,\label{eq.Esol}\\
{\bm\nabla}\times {\bf E}_{i} & = 0\label{eq.Eirr}.
\end{align}
\end{subequations}
In the presence of an isotropic medium, the $n$-th modes satisfy the Helmholtz equations
\begin{subequations}
\begin{align}
({\bm\nabla}^2 + \epsilon\, \omega^2_{s,n})  {\bf E}_{s,n} & = 0,\label{eq:helmholtz}\\
\epsilon\,\omega^2_{i,n} {\bf E}_{i,n} & = 0,
\end{align}
\end{subequations}
and the orthogonality conditions
\begin{subequations}
\begin{align}
\int dV {\bf E}_{s,n}({\bf x})\cdot {\bf E}_{s,m}^*({\bf x}) & = \delta_{n,m} \int dV |{\bf E}_{s,n}({\bf x})|^2,\label{eq:normfull}\\
\int dV {\bf E}_{i,n}({\bf x})\cdot {\bf E}_{i,m}^*({\bf x}) & = \delta_{n,m} \int dV |{\bf E}_{i,n}({\bf x})|^2,\\
\int dV {\bf E}_{s,n}({\bf x})\cdot {\bf E}_{i,m}^*({\bf x}) & = 0,
\end{align}
\end{subequations}
where $V$ is the volume of the PH. By Ohm's law, the current ${\bf j}_f$ is proportional to the E-field inside the system.
If the latter is expanded in terms of the PH modes, one finds
\begin{align}
{\bf E}({\bf x},t) & = \sum_n \left[ e_{s,n}(t) {\bf E}_{s,n}({\bf x}) + e_{i,n}(t) {\bf E}_{i,n}({\bf x})  \right],\label{eq.E}\\
{\bf j}_f({\bf x},t) & = \sum_n \left[ \frac{\omega_{s,n}}{Q_{s,n}} e_{s,n}(t) {\bf E}_{s,n}({\bf x}) + \frac{\omega_{i,n}}{Q_{i,n}} e_{i,n}(t) {\bf E}_{i,n}({\bf x})  \right],\label{eq.j}
\end{align}
where $Q_n$ and $\omega_{n}$ corresponds to the quality factor and the frequency of the $n$-th mode, respectively. The quality factor encodes the power $P$ lost from the system with stored energy $U$ per cycle, $Q\equiv P/\omega U$. The power lost, and in turn the quality factor, is traditionally split into two components, the power lost to resistive damping and the power extracted by the antenna, i.e., the power extracted from the cavity as the signal. The former gives the ``unloaded" quality factor, which considers only resistive losses, and the latter gives a combined ``loaded" quality factor which includes the signal extracted from the system.
For our purposes we will combine all losses into a single quality factor as in Eq.~\eqref{eq.j}. Note that one can also include the losses as part of dielectric constant, or a combination of $\epsilon$ and ${\bf j}_f$, however all of these prescriptions are equivalent.

With these equations, and assuming that the irrotational and solenoidal modes are not degenerate in frequency, we have all the ingredients we need to solve the main Eq.~\eqref{eq.main_eq_0}. We can do a Fourier transform in time, but leave explicit the spatial dependence, to get
\be
{\bm\nabla} \times {\bm\nabla} \times {\bf E} - \omega^2 {\bf D} =  i\omega {\bf j}_f + i\omega {\bf j}_{\rm eff}.
\label{eq.main_eq}
\ee 
By expanding into solenoidal and irrotational modes we can then solve for the E-fields that are excited using the Helmholtz equations. For an isotropic medium with $\epsilon_{ij}=\epsilon \delta_{ij}$ we find
\begin{subequations}
\begin{align}
&\left[\epsilon \omega^2_{s,n}-\epsilon \omega^2 - i\omega \frac{\omega_{s,n}}{Q_{s,n}} \right]  e_{s,n}  = \frac{i\omega \int dV {\bf j}_{\rm eff}\cdot {\bf E}^*_{s,n}}{\int dV |{\bf E}_{s,n}|^2},\label{eq:modesol}
\\
&-\left[\epsilon \omega^2 + i\omega \frac{\omega_{i,n}}{Q_{i,n}} \right]  e_{i,n}  = \frac{i\omega \int dV {\bf j}_{\rm eff}\cdot {\bf E}^*_{i,n}}{\int dV |{\bf E}_{i,n}|^2}, \label{eq:modeirr}
\end{align}
\end{subequations}
where we have neglected spatial dispersion, i.e. a spatial dependence in $\epsilon$. In practice, there is always some amount of spatial dispersion~\cite{Belov:2003,Demetriadou_2008}, but our main focus will be on modes for which the spatial derivatives are much smaller than the time derivatives (i.e. $\omega\gg k$). In this case the modification to $\epsilon$ due to spatial dispersion will be small and so safely neglected at the level of our analysis.
Writing ${\bf j}_{\rm eff}\equiv {\bf j}_{\rm eff}({\bf x})e^{-i\omega_g t}$ allows us to write the signal E-field induced by a GW at frequency $\omega_g$ as
\begin{align}
    {\bf E}({\bf x},t)&=-i\omega_g e^{-i\omega_gt}\nonumber\\
&\times\sum_n\left[ \frac{\eta_{s,n}{\bf E}_{s,n}({\bf x})}{\epsilon(\omega_g^2-\omega_{s,n}^2)+i\omega_g\frac{\omega_{s,n}}{Q_{s,n}}} + \frac{\eta_{i,n}{\bf E}_{i,n}({\bf x})}{\epsilon\omega_g^2+i\omega_g\frac{\omega_{i,n}}{Q_{i,n}}} \right]
\label{eq.Esignal}
\end{align}
where we have defined the overlap integral as
\be
\eta_{\alpha,n}=\frac{\int dV {\bf j}_{\rm eff}({\bf x})\cdot {\bf E}_{\alpha,n}^*({\bf x})}{\int dV |{\bf E}_{\alpha,n}({\bf x})|^2},
\label{eq.N}
\ee
and $\alpha= s, i$ indicates either solenoidal or irrotational modes. For maximally anisotropic media we only need to worry about the solenoidal modes, though the general case is mathematically more complex. Fortunately, for isotropic or maximally anisotropic media we can still use Eq.~\eqref{eq.Esignal}, as discussed in Appendix~\ref{appanio}.
From here on we will suppress the subscripts on $\eta$, as in practice we will be considering situations where there is only a single mode being excited at a given frequency in a single configuration.

We observe that, in principle, there are two types of resonances for the solenoidal and irrotational modes, in contrast to the case of an empty cavity where only the solenoidal modes resonate~\cite{Berlin:2021txa}. The former occurs when $\omega_g^2 = \omega_{s,n}^2$, while the latter happens when the real part of $\epsilon$ vanishes. Although it might seem from Eq.~\eqref{eq.Esignal} that the solenoidal modes would also resonate at $\epsilon = 0$, this condition is never met as long as a single-cavity mode formalism is applicable. This is because the Helmholtz equation for the solenoidal modes, Eq.~\eqref{eq:helmholtz}, requires $\epsilon > 0$.

\section{Medium Effects}\label{sec:mediumeffects}

Before analyzing the resonant conditions in Eq.~\eqref{eq.Esignal} and computing the signal power from the GW and PH interaction, we need to revise the mode frequencies in the presence of the dielectric medium and we need to study in detail the integral in Eq.~\eqref{eq.N}.

\begin{figure}[t]
\centering
\includegraphics[width=0.45\textwidth]{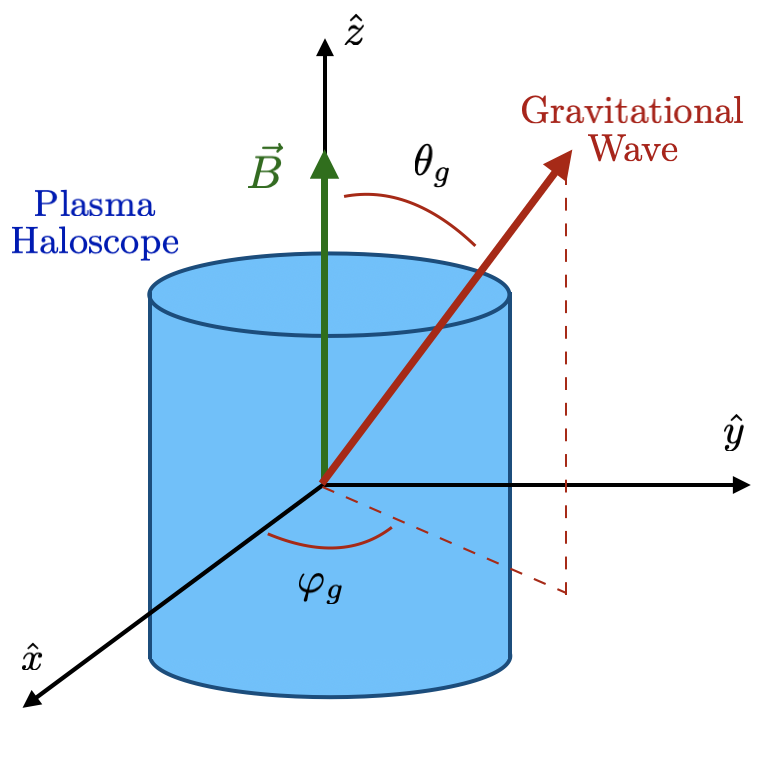}
\caption{Coordinates of the GW in the frame of the PH. The external B-field is aligned with the axis of the PH, and the GW passes through in the direction defined by the polar and azimuthal angles $\theta_g,\varphi_g$.}
\label{fg.angles}
\end{figure}

\subsection{Mode Frequencies}

Plasma haloscopes fall into a more general category of meta-materials, or more specifically, wire meta-materials \cite{PhysRevB.79.035118,PhysRevB.80.245101,PhysRevB.96.195132,Chen2018-mu,PhysRevB.104.L100304,SAKHNO2023101150}. While the explicit form of $\epsilon$ depends on the exact meta-material, for simplicity we will neglect spatial dispersion and assume that the permittivity of the system can be modeled as in a Drude metal,
\be
\epsilon(\omega)=1-\frac{\omega_{\rm p}^2}{\omega^2}, \label{eq:drude}
\ee
where the plasma frequency for a rectangular wire array with side lengths $a$ and $b$ is given by~\cite{Belov:2003} 
\begin{equation}
\label{eq:plasmafreq}
    \omega_{\rm p}^2 = \dfrac{2\pi/s^2}{\log\left(\dfrac{s}{2\pi r} \right)+F(u)}~, 
\end{equation}
where $s=\sqrt{ab}$, $u=a/b$ and
\begin{equation}
    F(u)=-\frac12 \log u+\sum_{n=1}^\infty\left(\frac{\coth(\pi n u)-1}{n} \right)+\frac{\pi u}{6} \, ,
\end{equation}
allowing for $\omega_{\rm p}/2\pi \sim$ GHz when $s \sim$ cm spacing. Here we have neglected losses as, in the single moded regime, they can be included in the overall quality factor.

When the plasma is inside a cavity and a single mode calculation is appropriate, we can read off the mode frequencies by inserting Eq.~\eqref{eq:drude} into Eq.~\eqref{eq.Esignal} while neglecting the imaginary component of $\epsilon$.

For the solenoidal modes in an isotropic medium we have
\begin{align}
\omega_{s,n}^2&=\omega_{\rm p}^2+k_{s,n}^2\,,
\end{align}
where $k_{s,n}$ is the wavenumber associated with the cavity mode (it is the frequency of an empty cavity) and can be obtained from $k_{s,n}^2{\bf E}_{s,n}\equiv-\nabla^2{\bf E}_{s,n}$. If the medium is anisotropic with a plasma frequency only in the $z$ direction, we instead would get 
\begin{equation}
\omega_{n}^2=\frac{1}{2}(\omega_{\rm p}^2+k^2_{n}+\sqrt{(k_n^2+\omega_{\rm p}^2)^2-4k_{z,n}^2\omega_{\rm p}^2}),
\end{equation}
where here $k^2_n E_{z,n}\equiv -\nabla^2E_{z,n}$ and $k_{z,n}$ can be obtained from $k_{z,n}^2 E_{z,n}\equiv-\partial_z^2 E_{z,n}$. As the irrotational modes are non-dynamical for anisotropic media we only need to worry about the solenoidal modes and so suppressed the subscript $s$.

For the irrotational modes, the resonant frequencies are simply given by zeros of the real part of the permittivity. In deriving Eqs.~\eqref{eq:modesol}~and~\eqref{eq:modeirr} we assumed no spatial dispersion. In this limit all irrotational modes have the same frequency $\omega=\omega_{\rm p}$, which would lead to high levels of mode mixing.
Mixed modes are very difficult to use for new physics searches as they have non-trivial combinations of the modes that form them. Without detailed characterization at every frequency one cannot know how such modes couple to gravity and they cannot be used for measurement.
As we discussed earlier, in practice, wired systems include some spatial dispersion,
so $\epsilon(\omega)\to \epsilon(\omega,k)$, 
which can allow for the irrotational mode to be separated enough for their categorization and use for detection. As we are mostly concerned with modes with $\omega\gg k$ even a relatively large spatial dispersion will have a small effect on the modes, primarily leading to just a small shift in the resonant frequency without changing the spatial properties of the mode (i.e., the overlap integral). Because of this we will use Eq.~\eqref{eq:modeirr} and assume that the mode frequencies are separated by at least the line width $\omega/Q$, allowing for a single mode to be selected.

\begin{figure}
\includegraphics[width=.47\textwidth]{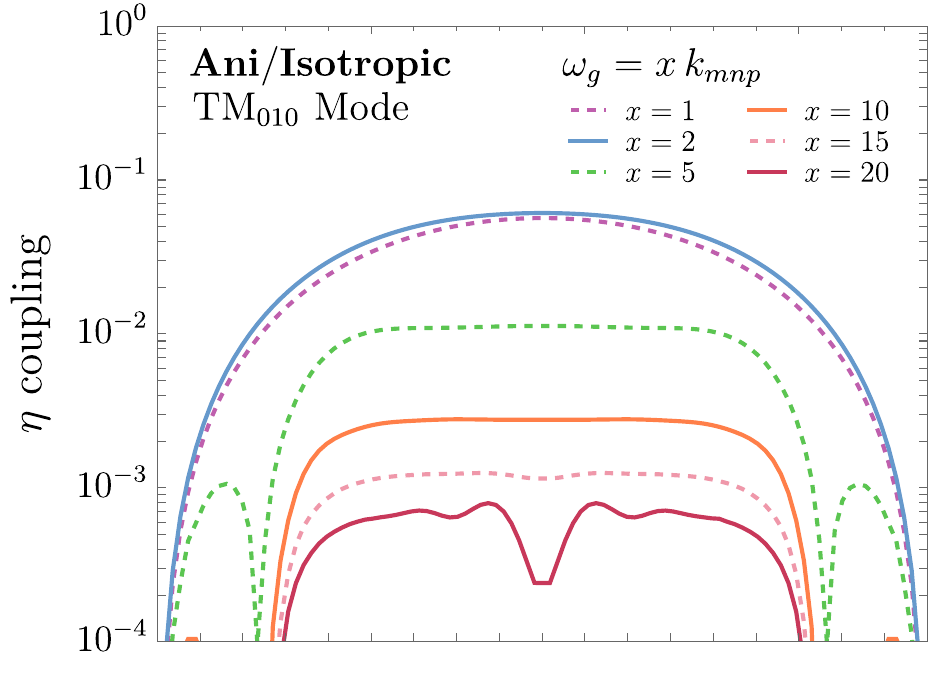}
\includegraphics[width=.47\textwidth]{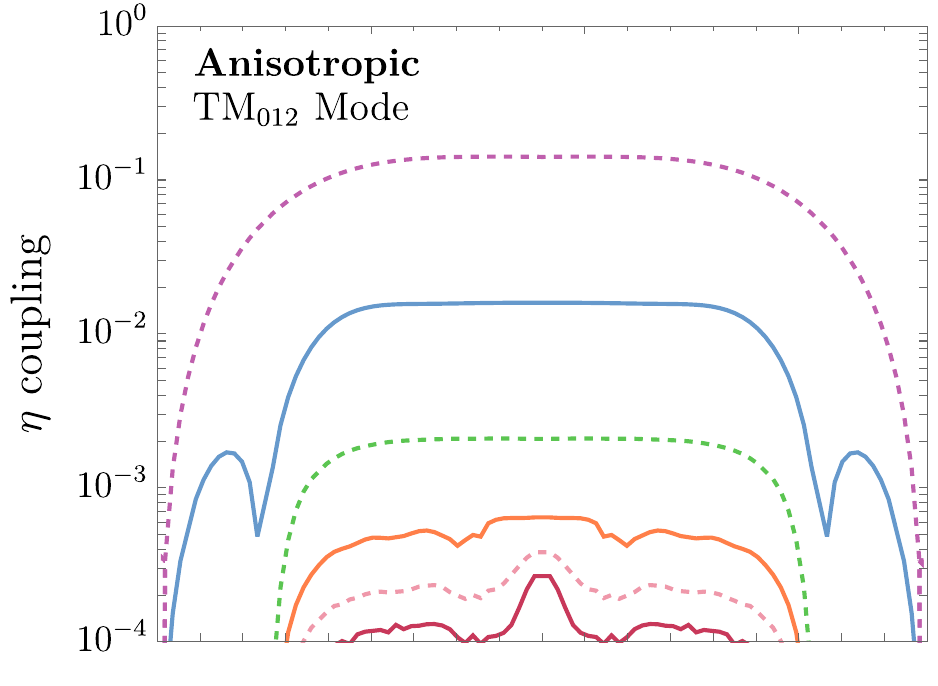}
\includegraphics[width=.47\textwidth]{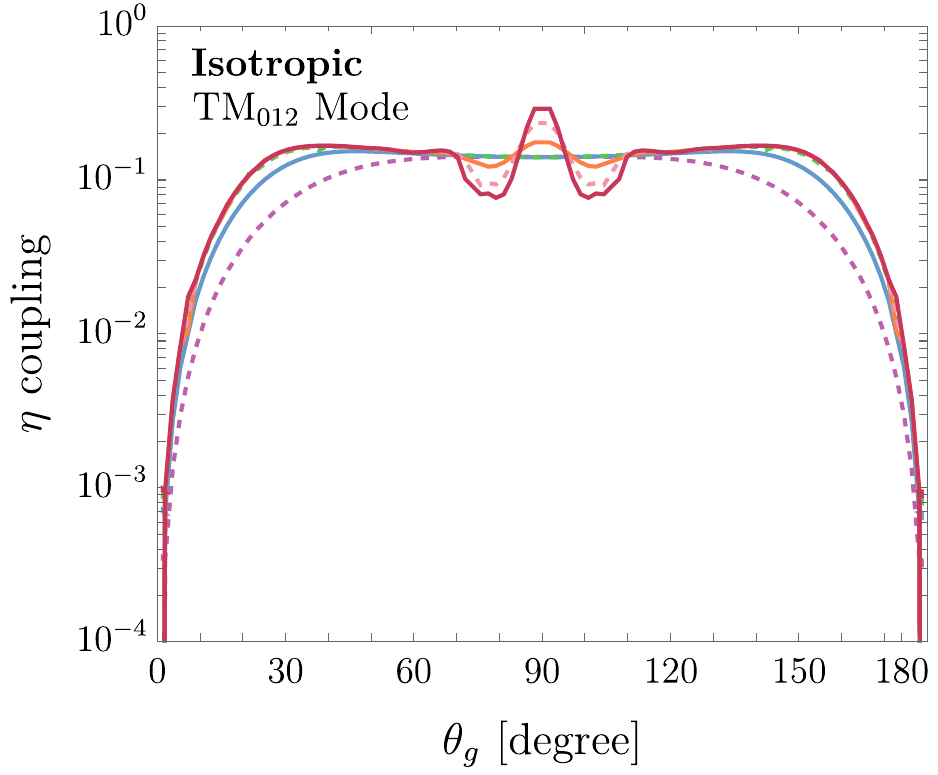}
\caption{Dimensionless coupling $\eta$ as a function of the GW polar angle $\theta_g$ for the mode $\rm TM_{010}$ (top), and the mode $\rm TM_{012}$ with an anisotropic (center) and an isotropic (bottom) plasma. Here, $k_{mnp}$ corresponds to the frequencies of the empty cavity, where the indices $mnp$ match those of the particular mode in each panel. The GW frequency is equal to that of the corresponding mode. The results in all panels correspond to a GW with $h_\times$ polarization, while the $h_+$ polarization provides negligible coupling.
}
\label{fg.TM_modes}
\end{figure}

\begin{figure*}
\centering
\includegraphics[width=0.360
\textwidth]{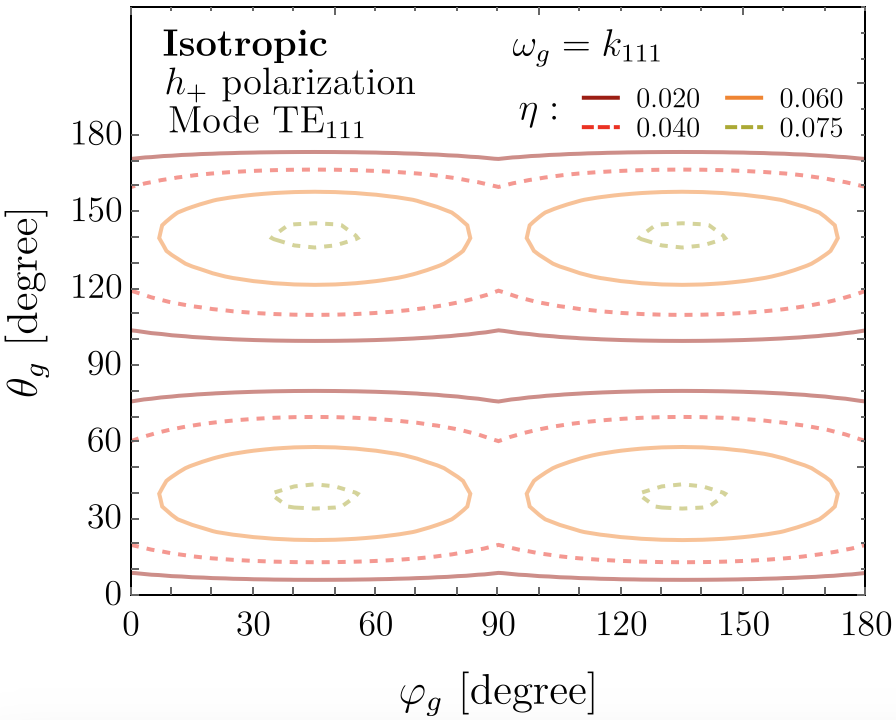}
\includegraphics[width=0.311\textwidth]{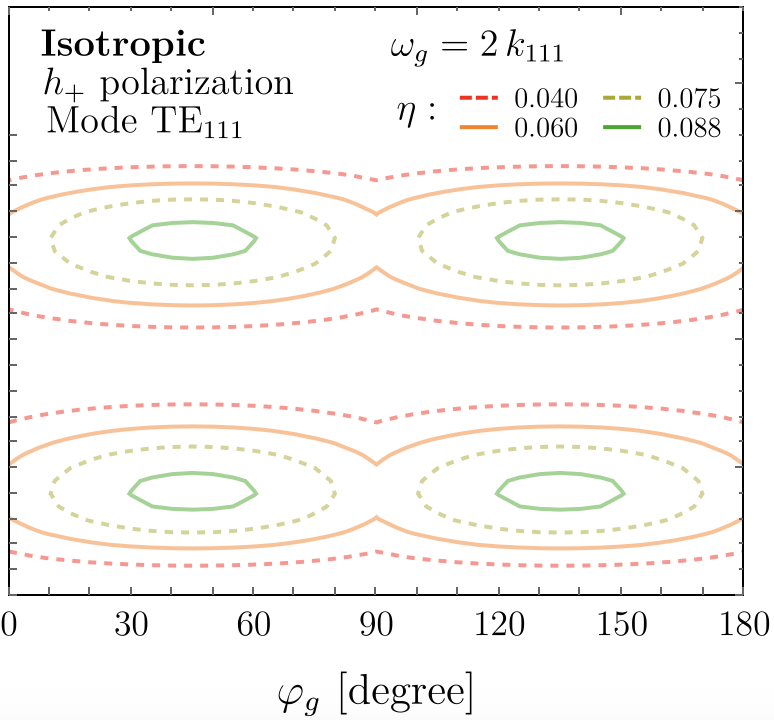}
\includegraphics[width=0.311\textwidth]{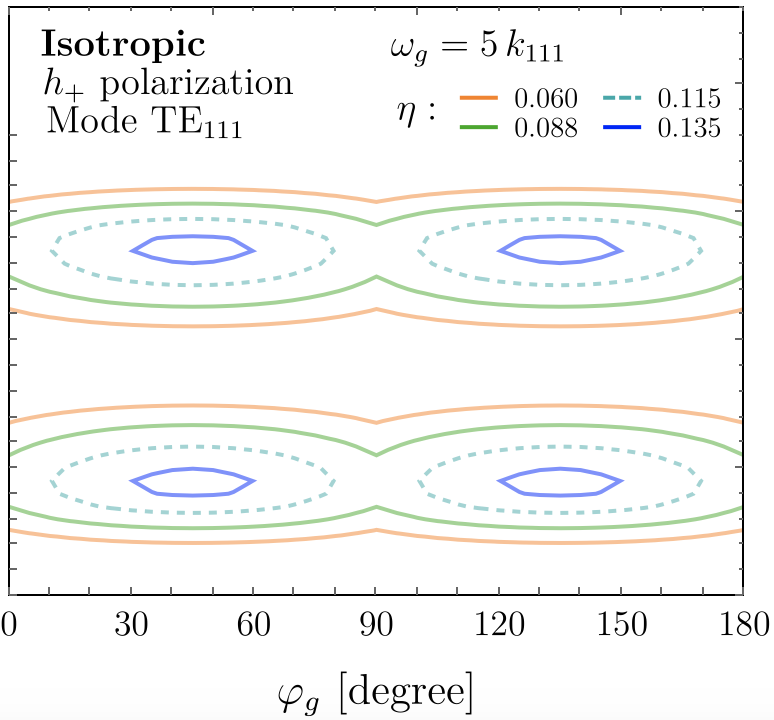}
\caption{Contours of the $\eta$ coupling as a function of the GW angles for the mode $\rm TE_{111}$ in a cylindrical cavity. The evident azimuthal variation comes from the $\varphi$ dependence of the mode. As the frequency increases, the coupling increases as well as one can see by comparing the contours in the left panel (low frequency $\omega_g=k_{111}$, small couplings $\eta=0.02$) with those of the right panel (high frequency $\omega_g=5k_{111}$, large couplings $\eta=0.1$). For this particular example we show the $h_+$ polarization only, but we verified that $h_\times$ provides similar results.
}
\label{fg.TE111}
\end{figure*}

\begin{figure*}
\centering
\includegraphics[width=0.360
\textwidth]{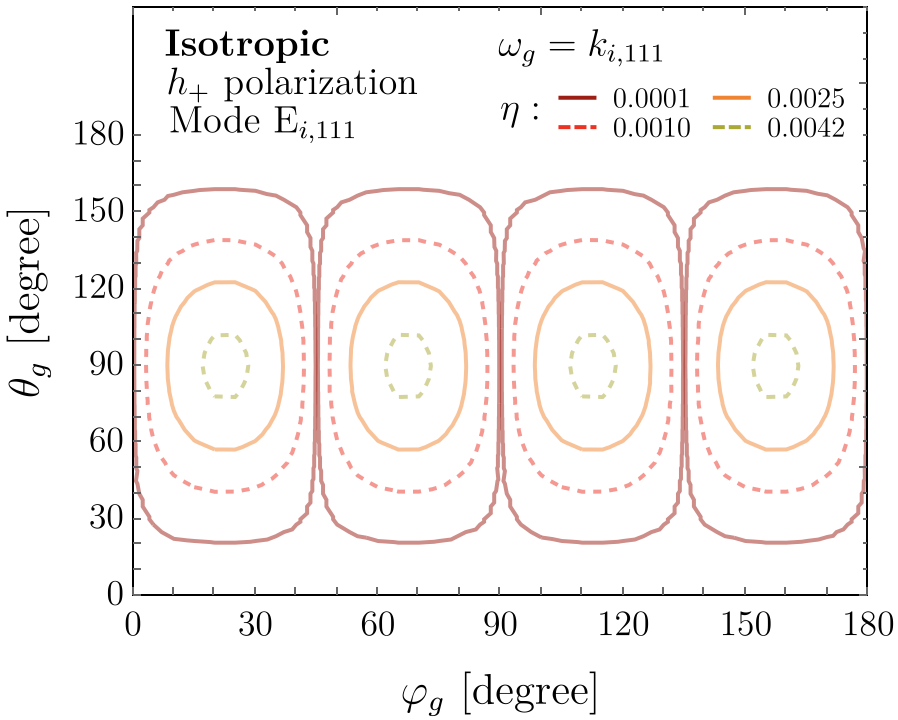}
\includegraphics[width=0.311\textwidth]{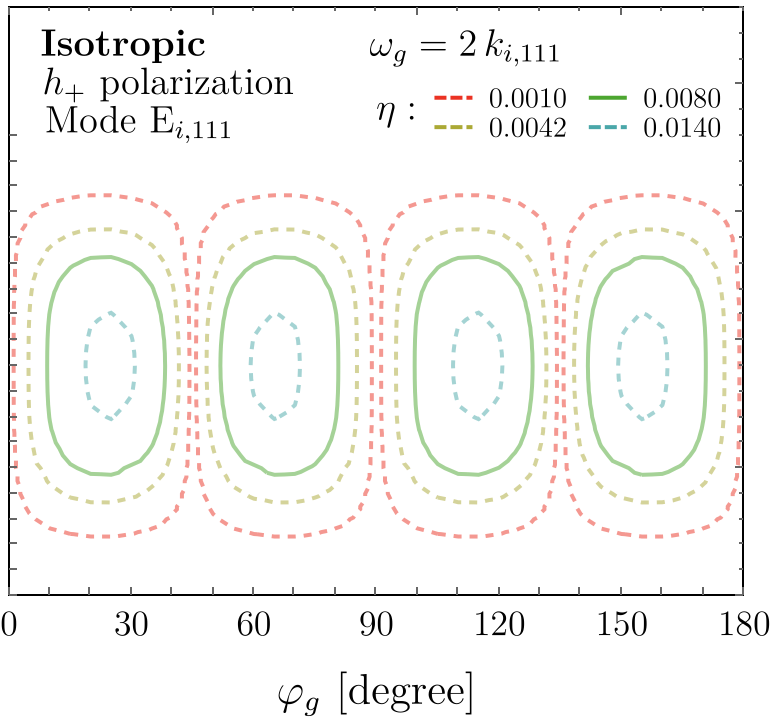}
\includegraphics[width=0.311\textwidth]{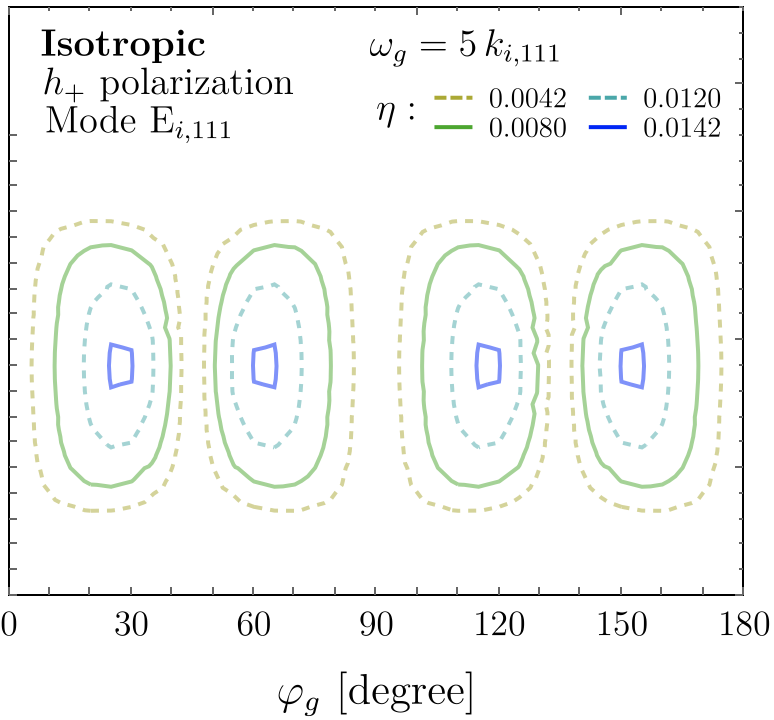}
\caption{Similar to Fig.~\ref{fg.TE111} but for the irrotational mode $E_{i,111}$ in a cubic cavity. Likewise, both polarizations $h_{+,\times}$ provide similar results.}
\label{fg.Ei111}
\end{figure*}

\subsection{Overlap Integral}
Following the discussion in Ref.~\cite{Berlin:2021txa}, from Eq.~\eqref{eq.N} we define the dimensionless coupling $\eta_{\alpha,n}^{\lambda}=|\eta_{\alpha,n}|/(B_0\omega_g^2V^{5/6}h_\lambda)$, 
where the index $\lambda=
(+,\times)$ represent the polarizations of the GW, and the currents $\hat{\bf j}_{\lambda}$ are obtained from
\be
{\bf j}_{\rm eff}({\bf x}) = B_0 \omega_g^2 V^{1/3} [h_+ \hat{\bf j}_+({\bf x}) + h_\times \hat{\bf j}_\times({\bf x})].
\ee
The coupling $\eta$ is an important parameter that determines the strength with which the GW excites a particular mode of the PH. From Eq.~\eqref{eq.Esignal}, we see that when the frequency of the GW matches the frequency of the $n$-th mode, the signal E-field is maximized. However, if the coupling $\eta$ vanishes for that particular mode, the signal power in the PH is null. Though it is not immediately clear from its definition, the coupling $\eta$ depends on the direction of propagation of the GW as well as its frequency. To describe this, we first define the coordinate system as shown in Fig.~\ref{fg.angles}. The external B-field is fixed along the axis of the PH, which in turn is defined to be along the $z$-axis. Finally, the angles $(\varphi_g, \theta_g)$ define the direction of the GW.

We now compute the $\eta$ coupling for three solenoidal: $\text{TM}_{010}$, $\text{TM}_{012}$, $\text{TE}_{111}$, and one irrotational: $\text{E}_{i,111}$ modes. The notation and properties of these modes is explained in detail in Appendices~\ref{app.modes}~and~\ref{app.irr}. 
These modes are chosen simply because they are the lowest frequency of each type of mode in the PH with non-negligible coupling to GWs, with the addition of the $\text{TM}_{012}$ mode which has been advocated as having a particularly high coupling~\cite{Berlin:2021txa,Gatti:2024mde}.
The fundamental frequencies of the solenoidal modes for the case of an empty cavity, are denoted by $k_{010}$, $k_{012}$, and $k_{111}$, respectively. In the case of the irrotational mode, for an empty cavity the frequency is zero, i.e., the mode is non-dynamical. However, we can still express the frequency in units of
the wavenumber $k_{i,111}$. In all cases, we assume that the frequency of the incoming GW matches that of the particular mode of interest.
As discussed earlier, the presence of the plasma allows us to scan over frequencies above that of an empty cavity by increasing $\omega_{\text{p}}$. To analyze the coupling $\eta$ at high frequencies, we define the coefficient $x = \omega_{\text{mode}}/k_{\text{mode}}$ and quantify the behavior of $\eta$ as a function of $x$.

\begin{figure*}[t]
\centering
\includegraphics[width=0.47\textwidth]{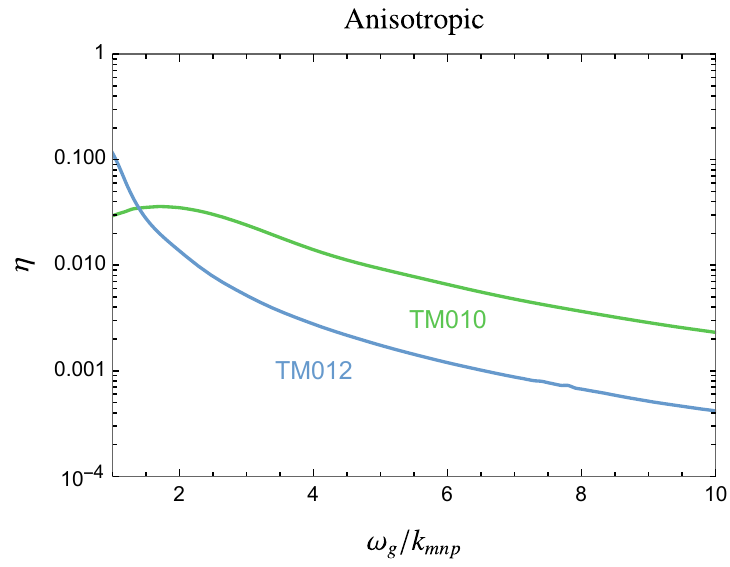}~~~\includegraphics[width=0.47\textwidth]{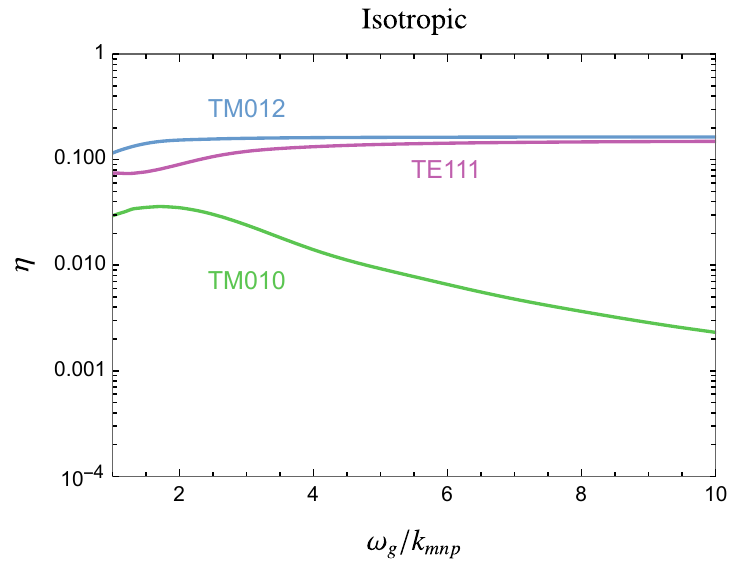}
\caption{Dimensionless coupling $\eta$ as a function of frequency for an anisotropic (left) and isotropic (right) plasma. We plot the TM$_{010}$, TM$_{012}$ and TE$_{111}$ modes, though the TE$_{111}$ is only shown for isotropic media as for the anisotropic case it simply matches the empty cavity.  Here, $k_{mnp}$ corresponds to the frequencies of the empty cavity, where the indices $mnp$ match those of the particular mode plotted. The GW frequency is equal to that of the corresponding mode. The results in all panels correspond to a GW with $h_\times$ polarization, while the $h_+$ polarization provides negligible coupling. Both angles $\theta_g,\phi_g$ of the GW angles are fixed to 45 degrees. }
\label{fg.HF}
\end{figure*}

\begin{itemize}
    \item Transverse Magnetic Mode $\rm TM_{010}$:
\end{itemize}

Figure~\ref{fg.TM_modes} (top) shows the $\eta$ coupling for the $\text{TM}_{010}$ mode as a function of the GW polar angle $\theta_g$. The azimuthal symmetry of the system and of that the mode makes $\eta$ independent of $\varphi_g$.
As discussed earlier, the anisotropic case refers to a PH with wires only in the $z$ direction, corresponding to the baseline setup for the ALPHA experiment~\cite{Lawson:2019brd,ALPHA:2022rxj}.
Only the $\times$ polarization of the GW contributes to $\eta$; we verified that the $+$ polarization results in a highly suppressed coupling. From the figure we see how the $\eta$ coupling tends to zero when the GW propagates parallel or anti-parallel to the B-field and reaches its maximum at a polar angle of $90^\circ$.

We note that the $\text{TM}_{010}$ mode behaves the same whether the PH is isotropic or anisotropic. This is because this mode has electric field components only along the $z$ axis. Since the anisotropic medium has wires only along the $z$ axis, making the PH isotropic does not affect the field of the mode.

The dashed-purple lines in Fig.~\ref{fg.TM_modes} represent the case of an empty cavity, where $x=1$. 
For this particular mode,
as we increase the frequency of the mode and that of the GW (recall that $\omega_g = \omega_{\text{mode}}$), the GW couples less efficiently to the mode. Depending on the angle, increasing the mode frequency $\omega_{010}$ by a factor of 20 above $k_{010}$ (solid-red line) can decrease $\eta$ by about two orders of magnitude.

\begin{itemize}
    \item Transverse Magnetic Mode $\rm TM_{012}$:
\end{itemize}

Fig.~\ref{fg.TM_modes} shows the $\eta$ coupling for the $\text{TM}_{012}$ mode in the case of an anisotropic PH (center). Similar to the $\text{TM}_{010}$ mode, the $\text{TM}_{012}$ mode exhibits azimuthal symmetry and a negligible coupling when the GW propagates along the axis of the PH, with maximal coupling occurring at a polar angle of 90 degrees. In this figure, we have only considered the $\times$ polarization of the GW, as the $+$ polarization exhibits negligible coupling. As the frequency of the mode
increases, the $\text{TM}_{012}$ mode decouples, even faster than the $\text{TM}_{010}$ mode.

Unlike the $\text{TM}_{010}$ mode, the $\text{TM}_{012}$ mode has electric field components along all three axes. For this reason, changing the medium from anisotropic to isotropic can alter the system in a non-trivial way. This behavior is illustrated in Fig.~\ref{fg.TM_modes}~(bottom). Note that the coupling does not vanish at higher frequencies; on the contrary, for certain angles, the coupling appears to increase until it reaches a plateau. In this case the TM$_{012}$ and E$_{i,111}$ modes give very similar results.

\begin{itemize}
    \item Transverse Electric Mode $\rm TE_{111}$:
\end{itemize}

This mode does not exhibit azimuthal symmetry. For this reason, we show in Fig.~\ref{fg.TE111} contours of the $\eta$ coupling as a function of the angles $(\varphi_g, \theta_g)$. The signal is periodic, with a period of $90^\circ$ in both angular directions. The figure displays the coupling corresponding to the `+' GW polarization. The `×' polarization yields similar numerical values for the coupling but has a different angular dependence.
In Fig.~\ref{fg.TE111}, there appears to be a `hot spot' where the coupling is maximal, corresponding to the angles $(\varphi_g, \theta_g) \sim (45^\circ, 45^\circ)$ degrees. At this angular point, the coupling increases from $\eta \sim 0.075$ (dashed-green line, left panel) to $\eta \sim 0.147$ (solid-blue line, right panel) as the frequency rises from the empty cavity case $\omega_g=k_{111}$ to a frequency five times higher $\omega_g=5k_{111}$.

\begin{itemize}
    \item Irrotational Mode $\rm E_{i,111}$:
\end{itemize}

This mode behaves quite similarly to the $\text{TE}_{111}$ mode in that it does not exhibit azimuthal symmetry and that its coupling increases as the frequency increases (Fig.~\ref{fg.Ei111}). This is for the case of an isotropic PH. The `hot spot' seems to appear for the angular values $(\varphi_g, \theta_g)\sim(25^\circ, 90^\circ)$. 
As the frequency increases by a factor of five the coupling at the `hot spot' increases by about a factor of three (compare the dashed-green line in the left panel with the solid-blue line in the right panel). Although the coupling increases with frequency, we can see in the figure how the coupling for this mode is about one order of magnitude below that of solenoidal modes studied above. In spite of this small coupling, it is interesting to see how irrotational modes can be indeed excited, and can in principle generate a signal from GW. As the modes are only dynamical in a medium, we should note that $k_{i,111}$ is not the frequency of the empty cavity, which is zero, but simply a wavenumber.

The high-frequency behavior of our modes of interest is shown in Fig.~\ref{fg.HF}. By studying these four modes $\text{TM}_{010}$, $\text{TM}_{012}$, $\text{TE}_{111}$, and $\text{E}_{i,111}$, we can identify the following pattern: The coupling between the GW and the PH tends to decrease with increasing frequency in the case of anisotropic systems (Fig.~\ref{fg.HF},~left). This suggests that, in its current configuration, ALPHA will lose sensitivity at higher frequencies. However, by considering an isotropic setup, the coupling can actually increase as the frequency rises (Fig.~\ref{fg.HF},~right). This nontrivial behavior was not accounted for in previous studies that attempted to estimate the projected sensitivity of PH in their potential use as GW detectors. Although the mode $\rm TM_{010}$ seems to be an exception to this rule, this is just an artifact of our choice to fix the B-field along the axis of the system.
If we were to consider a misalignment between the B-field and the PH axis, the mode $\rm TM_{010}$ would have $x,y$ components, and those would couple to the GW with an increasing strength at higher frequencies. 
It is also worth noting that the mode $\rm TE_{111}$ does not have an anisotropic case. This is also due to our choice to fix B along the PH axis. Because TE modes only have E-field components perpendicular to the PH axis, in the fully anisotropic case they behave exactly as in a empty cavity with a fixed frequency.

At a first glance, the different scaling between the isotropic and anisotropic modes seems extreme, however it can be understood by looking at the non plane wave structure in the PD frame. For an example, we can look at ${\bf J}_{\rm eff}$ for the case where the B-field is in the $z$ direction. For simplicity, we take the incoming wave to arrive from the $x$-axis and take the cross polarization (i.e., $h=h_\times$), though this choice does not have any qualitative difference. In this case, ${\bf J}_{\rm eff}$ is given by
\be
{\bf J_{\rm eff}}=B_0h_\times\Bigg(-\frac{z (x \omega_g-2 i) c_1}{2 \sqrt{2} x^3 \omega_g}, -\frac{y z c_2}{\sqrt{2} x^4 \omega_g},
  \frac{c_3}{\sqrt{2} x}\Bigg),
\ee
where
\begin{align}
\nonumber
c_1&=\left(x^2 \omega_g^2+e^{i x \omega_g} (-2+2 i x \omega_g)+2\right),\\
\nonumber
c_2&=\left(x^3 \omega_g^3-i x^2 \omega_g^2+2 x \omega_g+e^{i x\omega_g} (4 x \omega_g+6 i)-6i\right),\\
\nonumber
c_3&=1+e^{i x \omega_g} (-1+i x \omega_g).
\end{align}
At high ratios of $\omega_g/k_{mnp}$ (large plasma frequencies), our device is much larger than the Compton wavelength. For $x,y,z\gg 1/\omega_g$ we can simplify the current to
\begin{equation}
  {  \bf J}_{\rm eff}\simeq B_0h_\times\left(\frac{ z\omega_g^2}{2 \sqrt{2}},-\frac{y
 z\omega_g^2 }{\sqrt{2} x},\frac{i \omega_g
   e^{i x \omega_g}}{\sqrt{2}}\right).
   \label{eq:aeta}
\end{equation}
While the $x$ and $y$ components grow linearly with the size of the detector, the $z$ component is oscillatory; when the overlap integral with a slowly varying mode is calculated the $z$ component will tend to average to zero. For a steady $\eta$ one must then rely on the transverse components $E_{x,y}$. As shown in Appendix~\ref{app.modes}, for TM modes with wires only in the $z$ direction the E-fields in the $x,y$ directions are suppressed by a factor of ${\cal O}(k_{mnp}^2/\omega_g^2)$, which corresponds to a factor of $\epsilon$. Conversely, for an isotropic medium the modes higher than $\text{TM}_{010}$ have transverse E-fields on the same order as $E_z$.

From this we can see that the geometry for the planned ALPHA experiment leads to $\eta$ decoupling at high frequencies (i.e., large plasma frequencies) regardless of the chosen mode. This can be mitigated by either making the medium isotropic (or at least, having a non-trivial $x$ or $y$ component) or by aligning the E-field in a different direction. While the latter is mechanically easier, it would remove the sensitivity to axion dark matter, which is the main goal of the experiment.

\begin{figure}[t]
\centering
\includegraphics[width=0.47\textwidth]{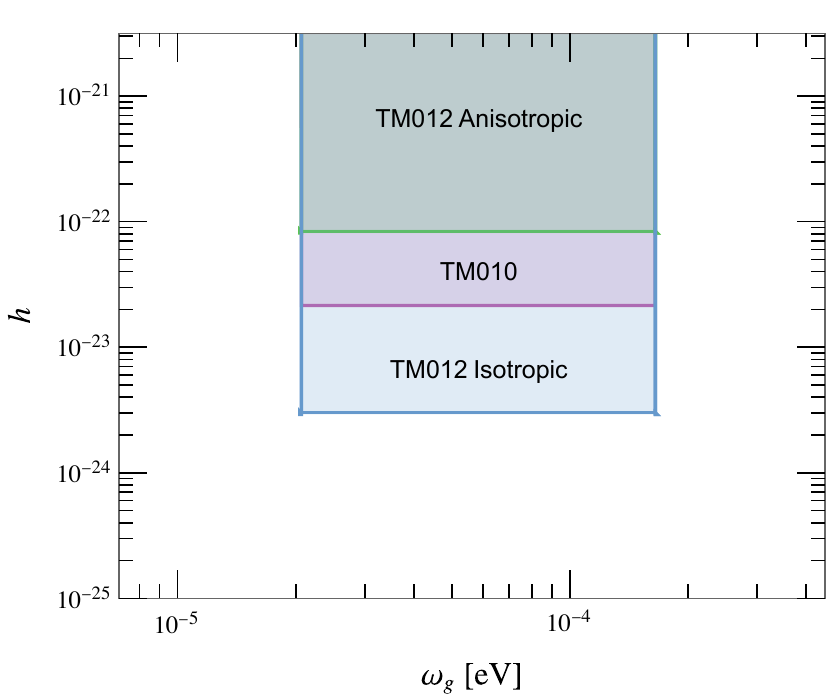}
\caption{Here we show the sensitivity of ALPHA to GWs, assuming a constant GW of linewidth $10^{-4}\omega$ with both angles $\theta_g,\phi_g$ fixed to 45 degrees. We take the experimental parameters from Ref.~\cite{ALPHA:2022rxj} assuming a cylindrical haloscope of 35~cm radius and  75~cm length inside a 13~T magnet. We assume that the experiment scans the range of $20-185~\mu$eV ($5-45$~GHz) with a quantum limited amplifier over two years. In blue we show the TM$_{012}$ mode in an isotropic media, with the anisotropic case shown in green. In between these the TM$_{010}$ mode is shown in lilac. }
\label{fg.proj}
\end{figure}

\section{Projections}\label{sec:projections}

Several sources have been considered to be the origin of high-frequency GWs. The sources can be described as either stemming from Early-Universe phenomena, which would result in a diffuse background, or from local events. As already mentioned in the introduction, the GW sources a PH could test are local. 

As pointed out in a seminal paper by Weinberg~\cite{Weinberg:1965nx}, any plasma can produce gravitons through bremsstrahlung (see also Ref.~\cite{Garcia-Cely:2024ujr} for additional processes). Therefore,
 even without accounting for novel physics, the Early-Universe plasma is expected to have copiously sourced GWs, whose amplitude today depends on the reheating temperature~\cite{Ghiglieri:2015nfa,Ghiglieri:2020mhm,Ringwald:2020ist,Ghiglieri:2022rfp}. Additional contributions might come from reheating itself (see e.g.~\cite{Ringwald:2020vei,Barman:2023ymn,Bernal:2023wus}), phase transitions (see e.g.~\cite{Witten:1984rs,Hogan:1986dsh,Hindmarsh:2013xza}) and from topological defects~\cite{Figueroa:2012kw,Figueroa:2020lvo} such as cosmic strings~\cite{Buchmuller:2019gfy,Servant:2023tua} or domain walls (see e.g. Figure 4 in Ref.~\cite{Gelmini:2020bqg}). The amplitude of GWs generated by these mechanisms is strongly constrained by $N_{\rm eff}$ measurements, so that any direct detection chance seems daunting at best. Some probes rely on GW-photon conversions over either large distances along the line of sight, e.g. from the Rayleigh-Jeans tail of the CMB ~\cite{Domcke:2020yzq} or from the diffuse photon flux converting in Earth, galactic, and extragalactic magnetic field~\cite{Ito:2023nkq,Lella:2024dus}, or on GW-photon conversion in the large fields of neutron-star magnetospheres~\cite{Ito:2023fcr,Dandoy:2024oqg,McDonald:2024nxj}. Unfortunately, these proposals fail to reach below the $N_{\rm eff}$ constraints, and have the additional drawback of missing the opportunity of detecting \textit{local} sources.

The background to searches of nearby exotic events is given by the unavoidable GW flux produced by stars, the closest being of course the Sun~\cite{Weinberg:1965nx,carmeli1967gravitational,1971reas.book.....Z,1985ApJ...288..789G}. However, physics beyond the standard model can outshine such background in GWs. A possible source of high-frequency signals is the annihilation of axions produced through superradiance~\cite{Arvanitaki:2012cn,Aggarwal:2020umq} which undergo $a+a\rightarrow \rm GW$ processes, the latter being possible by the presence of the BH which guarantees energy-momentum conservation. The most exciting prospect for the frequencies covered by a PH like ALPHA is probably the detection of PBHs~\cite{Dolgov:2011cq,Berlin:2021txa, Franciolini:2022htd} (see also Ref.~\cite{Profumo:2024okx} for a recent reappraisal accounting for an extended PBH mass function) or compact object~\cite{Giudice:2016zpa}  mergers, the latter being e.g. Q-balls, boson stars, oscillatons, or oscillons~\cite{Aggarwal:2020olq}, depending on the production mechanism. PBHs (or other compact objects) could have formed binaries, e.g., if produced close to each other in the Early Universe~\cite{Sasaki:2016jop}, or through dynamical capture in dense halos~\cite{1989ApJ...343..725Q,Mouri:2002mc}. Tantalizingly, PBHs can constitute 100\% of the dark matter for masses between $10^{-16}$ to $10^{-10}\,\rm M_\odot$~\cite{Carr:2020gox}, and the GWs emitted in their merger could be an important probe of their existence.
 
 While an exact sensitivity depends strongly on the assumed GW source, to allow for a comparison we consider a time constant GW signal of width $10^{-4}\omega$ and take the experimental parameters from Ref.~\cite{ALPHA:2022rxj}. This consists of a plasma haloscope with radius $35$~cm and length 75~cm with noise at the standard quantum limit inside a 13~T magnetic field. The experiment will scan between 5 and 45~GHz over a two year timeframe. To estimate the sensitivity we use Dicke's radiometery equation for the signal to noise ratio $S/N$,
 \begin{equation}
     \frac{S}{N}=\frac{P}{T_{\rm sys}}\sqrt{\frac{\Delta t}{\Delta \nu_B}},
 \end{equation}
where $T_{\rm sys}$ is the system temperature (given by $\omega$ at the standard quantum limit), $\Delta t$ is the measurement time and $\Delta \nu_B$ is the signal bandwidth. Assuming a benchmark $S/N=3$ and the above experimental parameters, we see the sensitivity to GW for ALPHA in Fig.~\ref{fg.proj}.

While an isotropic medium gives the clear best sensitivity with a higher order mode such as the TM$_{012}$ mode, the gain once one has integrated over a large frequency range is not as much as one might expect. Conversly, unlike what was suggested in Ref.~\cite{Gatti:2024mde} for an anisotropic medium the sensitivity is actually worse for the TM$_{012}$ mode than the base TM$_{010}$ mode, and about two orders of magnitude lower than suggested by Ref.~\cite{Gatti:2024mde}.

\section{Conclusions}\label{sec:conclusions}

The high-frequency range constitutes perhaps the last frontier of gravitational wave astronomy. While the chances of detecting a diffuse background seem currently faint, efforts based on the inverse Gertsenshtein effect could bring lucky observations of local events. Similarly to other axion detection schemes, such as cavities~\cite{Berlin:2021txa}, low-mass axion haloscopes~\cite{Domcke:2022rgu}, and dielectric haloscopes~\cite{Domcke:2024eti},
the  plasma haloscopes are also in principle sensitive to high-frequency gravitational waves.

While the  gravitational wave 
sensitivity of plasma haloscopes was briefly studied in Ref.~\cite{Gatti:2024mde}, several key aspects were neglected, leading to an overestimation of the sensitivity of an anisotropic plasma. In particular, the overlap integrals were taken to be the same as for an empty cavity, instead of depending strongly on the plasma frequency. As we have shown, for anisotropic wire media the overlap integral goes to zero
for large frequencies, leading to a two order of magnitude decrease in sensitivity to gravitational waves.

To avoid this shortcoming, we 
explored the behavior of isotropic media. Unlike in the anisotropic case, both solenoidal and irrotational dynamical modes are allowed. Further, as large E-fields are supported in all three directions the non-plane wave
components of gravitational waves 
allow for a nearly constant overlap integral regardless of frequency. 

Going forward, there are several directions that can be explored. Currently, the ALPHA experiment is designed around anisotropic media, and so loses sensitivity to gravitational waves, 
particularly at high frequencies. Tuneable designs with three dimensional wire arrays would recapture this sensitivity without impacting the search for axions. Further, the details of a gravitational wave signal depend strongly on the properties of the source. Detailed analysis techniques must be created to ensure that a signal is found regardless of the type of source. With these optimizations plasma haloscopes will be a powerful probe of high frequency gravitational waves. 

\acknowledgments

The authors thank Asher Berlin and Sebastian Ellis for useful discussions as well as Luca Visinelli, Claudio Gatti and Michael Zantedeschi for comments on the draft. AM and RC were supported
by the U.S. Department of Energy’s Office of Science under contract DE–AC02–76SF00515. AM was also supported by 
Fermilab’s
Superconducting Quantum Materials and Systems Center (SQMS) under contract number DE-AC02-07CH11359.
Fermilab is operated by the Fermi Research Alliance, LLC under Contract DE-AC02-07CH11359 with the U.S. Department of Energy. 
The work of GG and JH was supported in part by the U.S. Department of Energy (DOE) Grant No. DE-SC0009937.
EV is supported by the Italian MUR Departments of Excellence grant 2023-2027 ``Quantum Frontiers'' and by Istituto Nazionale di Fisica Nucleare (INFN) through the Theoretical Astroparticle Physics (TAsP) project.

\appendix
\section{Cavity Modes in Media}
\label{app.modes}
While the modes of a cylindrical cavity are well known, it is less common to include a medium such a plasma inside the cavity. To begin with, inside the cavity one can write down Maxwell's equations in the absence of free charges and currents
\begin{subequations}
\bea
{\bm\nabla}\cdot {\bf D} &=&0	,
\label{eq:Maxwell-a-matter2}\\
{\bm\nabla}{\bm \times} {\bf H} - \bf{\dot {D}}   &=& 0 ,
\label{eq:Maxwell-b-matter2}\\
{\bm\nabla}\cdot{\bf B}&=& 0\,,\label{eq:Maxwell-c2}\\
{\bm\nabla}\times{\bf E}+\bf{\dot{B}}&=&0.\label{eq:Maxwell-d2}
\label{eq:Maxwell-c-matter2}
\eea
\label{eq:Maxwell-x-matter2}
\end{subequations}
In general we wish to consider plasmas that have wires in either all three directions, or only in one direction aligned with the axis of the cylinder.
The latter would give a maximally anisotropic medium exhibiting a plasma frequency only for E-fields polarized in the direction
along the cylinder's axis. These cases are summarized in an electric permittivity tensor
\begin{equation}
{\bf D}=\epsilon{\bf E}=\begin{pmatrix}   \epsilon_t & 0 & 0  \\ 0 & \epsilon_t & 0\\ 0&  0&  \epsilon_z \end{pmatrix}{\bf E}\,
\end{equation}
where the transverse dielectric components are labled $\epsilon_t$ and the axial component 
$\epsilon_z$. We will consider the limits where $\epsilon_t=\epsilon_z(\epsilon_t=1)$ for the isotropic (anisotropic) cases. For simplicity we will consider a non-magnetic material with $\mu=1$. The symmetry of the system allows us to break up the E-fields into the transverse $t$ and z 
components 
\begin{equation}
		{\bf B}={\bf B}_t+B_z \hat {\bf z}\,;\quad  {\bf E}={\bf E}_t+E_z \hat {\bf z}.
\end{equation}
We can 
expand the E-fields into harmonic components, 
assuming that the E-fields oscillate with angular frequency $\omega$. We take the fields to be functions of $\omega, \vec{x}$, essentially Fourier transforming only the temporal parameters, to get
\begin{subequations}
\begin{align}
	({\bm\nabla}_t  +\frac{\partial}{\partial z}\hat{\bf z}){\bm \times} ({\bf B}_t+B_z \hat {\bf z}) & = -i\omega \(\epsilon_t{\bf E}_t+\epsilon_z E_z\hat {\bf z}
 \)
  ,\\
({\bm\nabla}_t   +\frac{\partial}{\partial z}\hat{\bf z}){\bm \times}({\bf E}_t+E_z\hat {\bf z} ) & =i \omega ({\bf B}_t+B_z \hat {\bf z}).
\end{align}
\end{subequations}

 Decomposing these equations into the z-direction and the transverse directions (the
transverse curl of the transverse vectors is necessarily in the z direction)
\begin{subequations}
\bea
\hat {\bf z}\cdot {\bf \nabla}_t\times {\bf B}_t&=&-i\omega\epsilon_z E_z, \\
\hat {\bf z}\cdot {\bf \nabla}_t\times {\bf E}_t&=&i\omega B_z ,\\
\hat {\bf z}\times  \frac{\partial {\bf B}_t}{\partial z}+{\bf \nabla}_tB_z\times \hat {\bf z}&=&-i\omega \epsilon_t{\bf E}_t, \\
\hat {\bf z}\times  \frac{\partial {\bf E}_t}{\partial z}+{\bf \nabla}_tE_z\times \hat {\bf z}&=&i\omega {\bf B}_t.
\eea
\end{subequations}
Taking the E-field $z$ dependence to be of the form $\cos{k_z z}$ or $\sin{k_z z}$, we get closed forms for ${\bf B}_t$ and ${\bf E}_t$;
\begin{subequations}
\bea
{\bf E}_t&=&\frac{1}{\epsilon_t\omega^2-k_z^2}\({\bf \nabla}_t\frac{\partial  E_z}{\partial z}+i\omega{\bf \nabla}_t{ B}_z\times \hat {\bf z}\),\\
{\bf B}_t&=&\frac{1}{\epsilon_t\omega^2-k_z^2}\({\bf \nabla}_t\frac{\partial  B_z}{\partial z}-i\omega\epsilon_t{\bf \nabla}_t{ E}_z\times \hat {\bf z}\).
\eea
\end{subequations}
Thus the transverse E-fields depend only on $E_z$ and $B_z$, corresponding to the TM and TE modes respectively. 

\subsection{Solenoidal modes}
For the axial components we can solve the following differential equations for Bessel functions 
\begin{subequations}
\bea
\frac{\epsilon_t}{\epsilon_t\omega^2-k_z^2}{\bf \nabla}^2_tE_z+\epsilon_zE_z &=&0\,\label{eq:helm} .\\
\frac{1}{\epsilon_t\omega^2-k_z^2}{\bf \nabla}^2_tB_z+B_z&=&0.
\eea
\end{subequations}
In cylindrical coordinates they become 
\begin{subequations}
\begin{align}
	\frac{\epsilon_t}{\epsilon_t\omega^2-k_z^2}\(r^2\frac{\partial^2 E_z}{\partial^2 r}+r\frac{\partial E_z}{\partial r}+\frac{\partial^2 E_z}{\partial^2\phi}\)+r^2\epsilon_zE_z=&\,0.\\
 \frac{1}{\epsilon_t\omega^2-k_z^2}\(r^2\frac{\partial^2 B_z}{\partial^2 r}+r\frac{\partial B_z}{\partial r}+\frac{\partial^2 B_z}{\partial^2\phi}\)+r^2B_z=&\,0.
\end{align}
\end{subequations}

    We can now write down the full mode structure for the TM and TE modes. We label the modes with subscripts $m,n,p$ to denote the number of spatial variations in each direction.  For the TM$_{mnp}$ mode we get
    \begin{widetext}
\begin{subequations}
\begin{eqnarray}
	E_r^\pm &=&- \left[\begin{pmatrix}A_+ \sin m\phi \\ A_-\cos m\phi\end{pmatrix}\right]\frac{ k_z}{\epsilon_t\omega^2-k_z^2}\sin(k_zz)J'_{m}\left(\frac{rx_{mn}}{R}\right)\frac{x_{mn}}{R},\label{eq:TM_r}\\
	E_\phi^\pm &=&-\left[\begin{pmatrix}A_+ \cos m\phi \\ -A_-\sin m\phi\end{pmatrix}\right]\frac{ k_z}{\epsilon_t\omega^2-k_z^2}\sin(k_zz)\frac{m}{r}J_{m}\left(\frac{rx_{mn}}{R}\right),\label{eq:TM_phi}\\
	E_z^\pm &=& \left[\begin{pmatrix}A_+ \sin m\phi \\ A_- \cos m\phi\end{pmatrix}\right]J_{m}\left(\frac{rx_{mn}}{R}\right)\cos(k_zz),\label{eq:TM_z}
\end{eqnarray}
\end{subequations}
\end{widetext}
where $J_m$ are the Bessel functions, $J_m'$ are the derivatives with respect to their argument, and $\phi$ is the azimuthal angle of cylindrical coordinates. Following the notation of Ref.~\cite{Berlin:2021txa} we use normalisation constants for the $\pm$ modes $A^\pm$ and define $x_{mn}$ as the $n$th zero of $J_m$. We use $R$ for the radius of the cavity. Note that the plasma changes the frequency at which these zeros occur, but not the overall shape of the Bessel functions. Whether or not the plasma is isotropic changes the relative weighting of the transverse components: for an isotropic plasma $\epsilon_t\omega^2=\omega_{mnp}^2$, where $\omega_{mnp}$ is the frequency of the corresponding mode of an empty cavity. In contrast, if $\epsilon_t=1$ and $\epsilon_z\ll 1$ then $\omega\gg k_z$, suppressing the transverse electric fields. 
    The TE modes similarly can be written down as
\begin{widetext}
\begin{subequations}
\begin{eqnarray}
    E_r^\pm&=&\left[{A_+ \cos m\phi \atop -A_- \sin m\phi}\right]\sin \left(k_zz\right)\frac{i \omega }{\epsilon_t\omega^2-k_z^2} \frac{m}{r} J_{m}\left(\frac{rx_{mn}'}{R}\right)\,,\label{eq:TE_r}\\
    E_\phi^\pm&=&-\left[{A_+ \sin m\phi \atop A_- \cos m\phi}\right]\frac{i\omega }{\epsilon_t\omega^2-k_z^2}\sin \left(k_zz\right) J_{m}'\left(\frac{rx_{mn}'}{R}\right)\frac{x_{mn}'}{R}\,,\label{eq:TE_phi}\\
    E_z^\pm&=&0,\label{eq:TE_z}
\end{eqnarray}
\end{subequations}
\end{widetext}
where $x_{mn}'$ is the $n$th zero of $J_m'$. For the anisotropic case we recover the modes of an empty cavity, as the $E_z$ component is trivial.

\subsection{Irrotational Modes }
\label{app.irr}

Maxwell's equations do not directly give the spatial structure of the irrotational modes, simply giving the resonant condition ${\rm Re}(\epsilon)=0$. However, inside some volume V within a surface S the irrotational modes can be 
derived from the scalar potential,
 defined as the orthogonal basis of solutions of the Helmholtz equation 
vanishing on the boundary~\cite{1124535},
\begin{subequations}
\begin{eqnarray}
(\nabla^2+k_n^2)\left.\phi_n \right|_{in\,V} &= 0 \\
\left. \phi_n \right|_{at\,S} &= 0.
\end{eqnarray}
\end{subequations}

Considering a cubic 
cavity of length $L$ and a coordinate system centered at the center of mass of the cavity the solutions to these equations are simply 
sinusoidal functions of the type $\sin\left(k_x[x-L/2]\right)$, where $k_x=n\pi/L$, $n=1,2,3...$ Thus the general solution is
\begin{multline}
\phi_{\ell mn}=A
\sin\left({\ell\pi\over L}\left[x-{L\over2}\right]\right)
\times\\
\sin\left({   m\pi\over L}\left[y-{L\over2}\right]\right)
\sin\left({   n\pi\over L}\left[z-{L\over2}\right]\right),
\end{multline}
where $A$ is a normalization constant.
Finally, the space profile of the irrotational modes is given simply by
\be
k_{\ell mn}{\bf E}_{i,\ell mn}({\bf x})=\nabla \phi_{\ell mn},
\ee
where $k_{\ell mn}$ is usually called a normalization factor and gives the wavenumber of the mode. Note that even for the lowest mode the E-field is non-trivial in all directions. Because of this, for an anisotropic medium where the dielectric constant is one in the $x,y$ directions a dynamical irrotational mode is not supported.

\section{Anisotropic Media}
\label{appanio}
In the main text, we wrote down the overlap integral assuming an isotropic medium. While the anisotropic case was addressed in Ref.~\cite{Gatti:2024mde}, there are some subtleties regarding mode normalization, propagated from a small error in Ref.~\cite{ALPHA:2022rxj}. As Ref.~\cite{ALPHA:2022rxj} was primary concerned with the TM$_{010}$ mode, which only has a $z$ component, rather than normalizing the modes via Eq.~\eqref{eq:normfull} they used
\begin{equation}
    \int dV  E_{z,n}({\bf x})  E_{z,m}^{*}({\bf x})  = V\delta_{n,m}.\label{eq:znorm}
\end{equation}
Aside 
 from an overall factor due to a different choice of amplitude, this condition only considers the $z$ components of the fields. To consider a general set of modes this condition does not guarantee that those modes are orthogonal. Unfortunately, this normalization condition was also used in Ref.~\cite{Gatti:2024mde}, and leads to errors when, for example, $\epsilon=1$. Here we will consider the maximally anisotropic case, with $\epsilon_t=1$ and $\epsilon_z\equiv \epsilon$. As the irrotational modes are non-dynamical in this case, we will suppress the $s,i$ subscript in this section and will only refer to solenoidal modes. Similarly, we will neglect the TE modes, as they are the same as for an empty cavity.

From Eq.~\eqref{eq.main_eq} we can split the fields into the transverse (noted with $t$) and axial (denoted with $z$) components,
\begin{equation}
    -\nabla^2{\bf E}-\omega^2({\bf E}_t+\epsilon {\bf E}_z)=i\omega{\bf j}_f+i\omega{\bf j}_{\rm eff}. 
\end{equation}
To proceed, we can use the free Helmholtz equations for each mode which give the dispersion relations
\begin{subequations}
   \begin{align}
\nabla^2E_{z,n}+\epsilon\omega_n^2 E_{z,n}=&0,\\
\nabla^2{\bf E}_{t,n}+\omega_n^2 {\bf E}_{t,n}=&0,
\end{align} 
\end{subequations}
and expand $E$ into modes given by
\begin{equation}
    {\bf E}=\sum_n e_n {\bf E}_n({\bf x}).
\end{equation}
Together these equations allow us to write
\begin{equation}
    \sum_n e_n (\omega_n^2-\omega^2)\left[{\bf E}_{t,n}+\epsilon{\bf E}_{z,n}\right ]=i\omega{\bf j}_f+i\omega{\bf j}_{\rm eff}.
\end{equation}

\begin{figure}[t]
\centering
\includegraphics[width=0.47\textwidth]{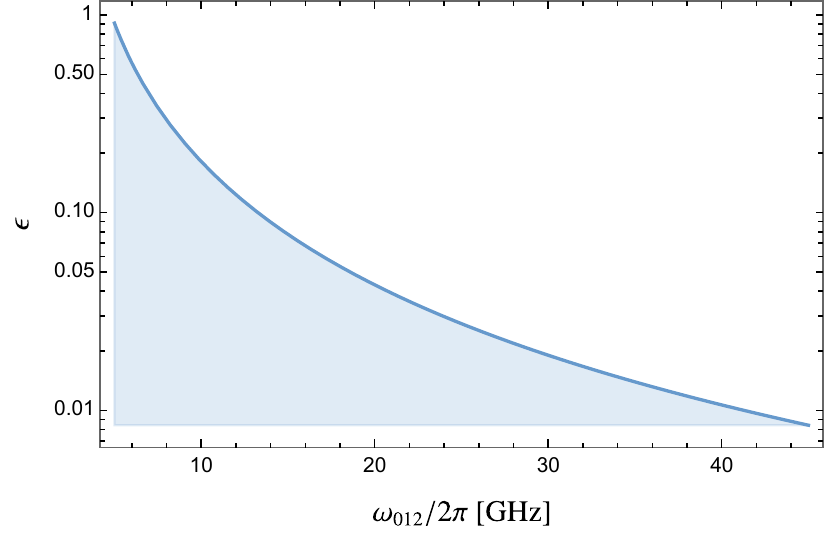}
\caption{Dielectric constant $\epsilon$ required for the TM$_{012}$ mode with resonant frequency $\omega_{012}$ in an anisotropic medium. We take the cavity to be cylindrical with a 35~cm radius and  75~cm length.  We assume that the experiment scans the range of ($5-45$~GHz) as shown in Fig.~\ref{fg.proj}. }
\label{fg.epsilon}
\end{figure}

    \begin{widetext}
In the isotropic case, it would be easy to project out onto a given mode using Eq.~\eqref{eq:normfull}, giving the usual overlap integral. In Ref.~\cite{Gatti:2024mde}, this equation was rewritten so that only $E_z$ appears, at the cost of additional terms on the right hand side involving ${\bf j}_{\rm eff}$ and using Eq.~\eqref{eq:znorm}. Regardless to proceed one must make an approximation. This can be quantified by noting that Eq.~\eqref{eq:normfull} implies that
\begin{equation}
    \int dV{\bf E}_{t,n}\cdot{\bf E}_{t,m}^*=\delta_{n,m} \int dV |{\bf E}_{n}|^2-\int dV{\bf E}_{z,n}\cdot{\bf E}_{z,m}^*,
\end{equation}
which can be combined with Eq.~\eqref{eq.j} to give
\begin{align}
\sum_n e_n (\omega_n^2-\omega^2)\left[\delta_{n,m}+(\epsilon-1)\frac{\int dV{\bf E}_{z,n}\cdot{\bf E}_{z,m}^*}{\int dV |{\bf E}_{m}|^2}\right]-i\omega\frac{\omega_n}{Q_n}e_n\delta_{n,m} =\frac{i\omega \int dV {\bf j}_{\rm eff}\cdot {\bf E}^*_{m}}{\int dV |{\bf E}_{m}|^2}.
\end{align}
\end{widetext}
This expression simplifies in two limits. The first is the trivial one, where $\epsilon=1$, which recaptures the expressions of Ref.~\cite{Berlin:2021txa}, a case for which the equations in Ref.~\cite{Gatti:2024mde} do not hold. The second is when $\epsilon\ll 1$. As discussed in Appendix~\ref{app.modes}, ${\bf E}_{t,n}={\cal O}(\epsilon{\bf E}_{z,n})$ so Eq.~\eqref{eq:znorm} is correct up to ${\cal O}(\epsilon^2)$ terms. In either case we can write
\begin{equation}
    {\bf E}({\bf x},t)\simeq-i\omega_g e^{-i\omega_gt}
\sum_n\frac{\eta_{n}{\bf E}_{n}({\bf x})}{\epsilon(\omega_{g}^2-\omega_n^2)+i\omega_g\frac{\omega_n}{Q_n}}\,
\end{equation}
 which is true exactly for $\epsilon=1$ and approximately for $\epsilon\ll 1$. The latter condition holds by design for almost every case considered in this paper. If $\epsilon\sim 1$, then there is little point to including a wire array, as the volume is almost the same as an equivalent frequency empty cavity. In this case we would have introduced more complexity for little gain. To show that our calculations are almost always in the region of validity, in Fig.~\ref{fg.epsilon} we show the $\epsilon$ required for a given frequency for the TM$_{012}$ mode over the $5-45$~GHz range considered in Fig.~\ref{fg.proj}. We see that $\epsilon\ll 1$ for all but the lowest frequencies, and so this approximation should not have a notable impact on the projected sensitivity.

\section{Signal Power}
\label{app.power}
Once explicit expressions for the generated E-field are found, the final ingredient to compute the reach of PH is to obtain the signal readout. The energy stored in an isotropic medium with temporal dispersion is given by
\begin{align}
    U=\frac{1}{4}\int\left ( \frac{\partial (\epsilon\omega)}{\partial\omega}|\mathbf{E}|^2+|\mathbf{B}|^2\right)dV\simeq\frac{1}{2}\int
    |     \mathbf{E}|^2 dV.\label{eq:U}
\end{align}
Note that the latter approximation holds even for $\epsilon\neq 1$, as the B-fields come from the spatial derivatives of the E-fields and so are smaller by a factor of $\sqrt{\epsilon}$. For the Drude model of $\epsilon$, the energy added by the magnetic field is opposite to the correction coming from $\partial (\epsilon\omega)/\partial \omega$, totaling to a factor of $1/2$.
To find the losses in the medium we can treat the resistive current in Eq.~\eqref{eq.j} as the imaginary component of the effective dielectric constant $\epsilon''\equiv\epsilon_z''=\epsilon_t''$, where $\epsilon''\equiv \omega_n/\omega Q_n$, using\footnote{Alternatively, one could also use Ohm's Law, which is equivalent.}~\cite{landau2013electrodynamics}
\begin{equation}
P=\frac{ \epsilon''}{2}\int \omega
\left|\mathbf{E}\right|^2 dV .
\label{eq:pwires}
\end{equation}
Substituting $\mathbf{E}$ with its mode expansion (Eq.~\eqref{eq.E})
and making use of the mode orthogonality conditions, we find 
\begin{equation}
    P=\frac{\epsilon''}{2}\sum_n 
    \omega
    \left[ |\tilde{e}_{s,n}(t)|^2 +  |\tilde{e}_{i,n}(t)|^2 \right]
\end{equation}
where 
\begin{subequations}
\begin{align}
    |\tilde{e}_{s,n}|^2&=|e_{s,n}|^2\int dV |\mathbf{E}_{s,n}(\mathbf{x})|^2 \\
|\tilde{e}_{i,n}|^2&=|e_{i,n}|^2\int dV |\mathbf{E}_{s,n}(\mathbf{x})|^2;
\end{align}
\end{subequations}
the mode amplitudes of the solenoidal and irrotational modes are obtained from Eqs.~\eqref{eq:modesol} and~\eqref{eq:modeirr}, respectively.
Assuming the GW to hit a resonance, $\omega=\omega_n$, we find
\begin{equation}
    P=\omega\frac{\epsilon''}{2}
  |\tilde{e}_{\rm res}(t)|^2 .
\end{equation}
Combining this result with Eq.~\eqref{eq.Esignal}, we see for a gravitational wave exciting a single resonance, the on resonance power is given by
\begin{equation}
    P=\frac 12 Q \omega_g^3V_{\rm cav}^{5/3} (\eta^\lambda h_\lambda B_0)^2,
\end{equation}
as in Ref.~\cite{Berlin:2021txa}.

\providecommand{\href}[2]{#2}\begingroup\raggedright\endgroup

\end{document}